\documentclass[aps,prb,amsfonts,amssymb]{revtex4}
\usepackage{amssymb}
\usepackage{amsmath}
\usepackage{graphicx}
\usepackage{citesort}

\def \be{\begin{equation}}
\def \ee{\end{equation}}
\def \ba{\begin{array}{l}}
\def \Ba{\begin{array}{ll}}
\def \ea{\end{array}}
\def \bq{\begin{eqnarray}}
\def \eq{\end{eqnarray}}
\def \nn{\nonumber\\}
\def \lb{\label}
\def \ln{{\rm ln}}

\def \fr{\frac}

\def \b{\beta}
\def \d{\delta}
\def \D{\Delta}

\def \f{\phi}

\def \lm{\lambda}

\def \n{\nabla}
\def \p{\varphi}
\def \h{\chi}

\def \t{\tau}

\def \tl{\tilde}
\def \ol{\overline}

\def \[{\left[}
\def \]{\right]}
\def \({\left(}
\def \){\right)}
\def \Tr{{\rm Tr}}
\def \R{R_{c}}
\def \T{T_{c}}

\def \I{\int d^{D}x}
\def \2{\frac{1}{2}}
\def \4{\frac{1}{4}}

\begin{document}

\title{Griffiths singularity in the random Ising ferromagnet}

\author{Vik.S.\ Dotsenko}

\affiliation{LPTMC, Universit\'e Paris VI, 75252 Paris, France} 
\affiliation{L.D.\ Landau Institute for Theoretical Physics, 
   117940 Moscow, Russia}

\date{\today}

\begin{abstract}
The explicit form of the Griffiths singularity in the random 
ferromagnetic Ising model in external magnetic field is derived.
In terms of the continuous random temperature Ginzburg-Landau 
Hamiltonian it is shown that in the paramagnetic phase 
away from the critical point the free 
energy as the function of the external magnetic 
field $h$ in the limit $h \to 0$ has the essential singularity 
of the form $\exp\[-(const)/h^{D/3}\]$ (where $1<D<4$ is the space
dimensionality). It is demonstrated that in terms of the replica 
formalism this contribution to the free energy comes due to 
off-perturbative replica instanton excitations.
\end{abstract}

\maketitle

\section{Introduction}

The history of the problem of the Griffiths singularities
starts from the theorem of Lee and Yang \cite{lee-yang}
which states that the partition function of an (ordered)
ferromagnetic Ising model (in any space dimensions and 
for any lattice connectivity)
as the function of the external magnetic field $h$ 
in the thermodynamic limit has a continuous distribution of 
zeros along the imaginary axis of the {\it complex} parameter $h = x + iy$.
Moreover, in the paramagnetic phase depending on the temperature
this distribution starts at finite distance $\D y$ from the real 
axis, which means that here the free energy of the 
system is an analytic function of the {\it real} magnetic 
field $h$ (with $y=0$). On the other hand, when the temperature $T$
approaches the phase transition point $\T$ from above the
value of the interval $\D y$ shrinks to zero, 
so that
the distribution of zeros touches the real axes 
at $T=\T$, 
which indicates (in agreement with the modern 
theory of the critical 
phenomena) that at the phase transition point 
the free energy must 
be a non-analytic function of (real) $h\to 0$.

Now, if one would try to apply
the above general observations for random systems, 
the consequences
turn out to be much more tricky. As an example, 
let us consider
bond diluted ferromagnetic Ising model, in which 
the critical
temperature $\T(p)$ is the (decreasing) function 
of the degree
of dilution $p$. Here one can note that in 
an infinite system
at a temperature $T$ which is above $\T(p)$ 
(such that the state 
of the system is paramagnetic) but below the 
critical temperature 
$T_{0}$ of the corresponding pure system, 
with a finite (exponentially small) 
probability there 
are exist arbitrary large less diluted 
ferromagnetic "islands" 
which are critical exactly at this given 
temperature $T$. 
It is important that
such clusters exist at any temperature in 
the interval
$\T(p) < T < T_{0}$. Thus one can expect that 
the free energy
of such random system must be a non-analytic 
function of the 
external magnetic field $h\to 0$ at {\it any} 
temperature between
$\T(p)$ and $T_{0}$, Ref[\onlinecite{grif}].
Unlike pure systems, however, here it is much 
more difficult
to predict the explicit form of such 
non-analyticity.

For the one-dimensional 
diluted Ising chain it has been shown that its 
free energy 
$F(T,h)$ is non-analytic in the point $T=0$, $h=0$,
 and moreover
the divergences of the pure system 
thermodynamics are replaced
by an essential singularity at which all 
functions are finite and
infinitely differentiable \cite{1D-wortis}.
According to further more general studies 
(in terms of heuristic 
arguments \cite{harris,imry} and the Bethe 
lattice systems \cite{harris}), the form of this 
singularity has been argued be
of the type $\exp\[ - (const)/h \]$.

At the same time there has been much interest 
in the dynamical properties of such systems
in the temperature interval $\T(p) < T < T_{0}$.
It has been discovered that the relaxation
processes here are slower than the exponential
\cite{grif-relax}. 
It turned out that due to the presence
of rare large ferromagnetic clusters
the relaxation of e.g. the order parameter
takes either
"stretched-exponential"
form $\exp\[- (t/\t)^{\b(T)}\]$ controlled
by the temperature dependent exponent $\b < 1$
(the result preferred by the numerical simulations
\cite{str-expon-numer} ), or even slower type of decay
$\exp\[- (const)(\ln t)^{D/(D-1)}\]$ (predicted
analytically \cite{grif-relax,grif-relax-cesi}).
Due to these quite non-trivial dynamical properties,
which are essentially different from the
paramagnetic phase, the state of the system 
in this temperature interval is usually called
the Griffiths phase.

Besides dynamics, an essential progress has been 
achieved in the analytical investigation of the
distribution of zeros of the partition function
alone the imaginary axis of the complex 
magnetic field. In particular, formal 
replica supersymmetric  calculations performed 
for the random temperature Ginzburg-Landau
Hamiltonian \cite{zeros-cardy} has demonstrated the
importance of the instanton-like off-perturbative 
excitations which provide the development
of the "tail" in the distribution of zeros
in analogy with the density of states 
in the Anderson localization problem.
On the other hand, the study of the diluted 
infinite-range Ising ferromagnet with 
{\it finite} connectivity 
in purely imaginary magnetic field $h = iy$
has shown that in the paramagnetic phase
the tail of the density of zeros $\rho(y)$
has the explicit form of the type
$\rho(y) \sim \exp\[ - f(T)/y \]$,
where the function of the temperature $f(T)$ 
vanishes in the critical point \cite{zeros-bray}.
Similar tail has been also
found numerically in the two-dimensional
diluted Ising model \cite{zeros-numer}.
All that, however, doesn't answer the main
question, what is the explicit form of the
Griffiths singularities of the thermodynamical
quantities as the function of the {\it real}
magnetic field in the point $h = 0$.

In this paper I am going to consider this problem
in terms of the $D$-dimensional random temperature 
Ginzburg-Landau Hamiltonian (in dimensions $1<D<4$)
in the paramagnetic phase away from the critical  
point. In the next section simple
heuristic arguments will be proposed which 
demonstrate on a qualitative level the physical 
mechanism by which non-analytic contributions 
appear in the thermodynamical functions
of such type of systems. After that, the systematic
method of off-perturbative replica calculations
will be formulated in Section III. Finally,
in Section IV it will be demonstrated that 
non-analytic (Griffiths) contributions to
the thermodynamics comes from off-perturbative
instanton-like excitations. In the limit $h \to 0$
such contributions to the free energy is argued to
have the explicit form
\be
\lb{gr1}
\D F \; \sim \; \exp\[ - C \; h^{-D/3} \]
\ee
where the constant $C$ is defined by the parameters
of the Ginzburg-Landau Hamiltonian and by the strength
of the disorder.

\section{Heuristic arguments}

Before starting doing systematic calculations let us try
to understand on a pure qualitative level,
using simple "hand-waving-arguments", what is 
the physical mechanism by which
non-analytic contributions are coming into the free
energy (and others thermodynamical functions)
in the paramagnetic phase of weakly disordered 
ferromagnetic Ising model in external magnetic field.
Let us suppose that such system in $D$ dimensions
can be described by the continuous Ginzburg-Landau 
Hamiltonian
\be
\lb{gr2}
H = \I \Biggl[ \2 \(\n\f(x)\)^{2} + 
     \2 \(\t - \d\t(x)\) \f^{2}(x) +  
  + \4 g \f^{4}(x) - h \f(x) \Biggr] 
\ee  
where the reduced temperature parameter $\t$ is taken to be
positive and sufficiently large to place the system
into the paramagnetic phase. The disorder is modeled by a
random function $\t(x)$ which is described by the 
Gaussian distribution,
\be
\lb{gr3}
P[\d\t] = p_{0} \exp \Biggl( -\frac{1}{4u}\I (\d\t(x))^{2} \Biggr) \ ,
\ee
where $u$ is the small parameter which describes 
the strength of the disorder,
and $p_{0}$ is the normalization constant. 

Intuitively, it is clear that non-trivial contributions to 
the thermodynamics are coming due to rare 
"ferromagnetic islands" in which the value
of $\d \t(x)$ is on average bigger than $\t$. Let us consider
such an island, 
which is characterized by the linear size $L$ and the typical
value of the "local temperature" $(\t - \d \t) = - \xi < 0$
Its probability is exponentially small,
\be
\lb{gr4}
{\cal P}\[L, \xi\] \; \sim \; 
\exp \Biggl( -\frac{(\t + \xi)^{2}}{4u} \; L^{D} \Biggr) \ ,
\ee
and therefore such islands are well separated from each
other and can be
considered as non-interacting.
Note that the physical machanism of
slowing down of the relaxational processes due to
the presence of these ferromagnetic islands are more or
less clear. The magnetization orientation  
of the ferromagnetic cluster can be either "up"or "down", 
and if its size $L$ is big, then it would require a long
"elementary relaxation time" $t(L)$ to flip it from 
one orientation to the other (since flipping 
of the cluster would require overcoming a big energy barrier,
which is proportional to the volume of the cluster $L^{D}$).
The origine of the non-analytic contributions
appearing in
terms of  pure statistical mechanics 
is much less clear: since time is formally infinite
here, the presence of big energy barriers, 
separating the two orientations is irrelevant.

\vspace{3mm}

First, let us consider what is going on in the zero 
external magnetic field. Here one could distinguish
two types of contributions to the thermodynamics:

(1) the perturbative one, coming from the spatial scales
smaller than the correlation length $R_{c}$,
which formally could be computed e.g. in terms of the
renormalization-group (RG) approach \cite{RG-disorder};

(2) the non-perturbative contributions (missing 
in the RG treatment) due to the 
"up" and "down" ferromagnetic states of rare
ferromagnetic islands
discussed above, which are coming from the spatial
scales bigger than the correlation length $\R$. 

It has to be noted that the island with small
(negative) local temperature $-\xi$ can be
characterized as having the distinct (mean-field)
"up" and "down" states only if its size is
much bigger than its local correlation length
$\R(\xi) \sim \xi^{-1/2}$  In what follows we are going to
use the mean-field values of the critical exponents
assuming that the temperature of the system 
is taken sufficiently far away from the critical 
region near $\T$ (the size of this region is 
of the order of $\t_{g} = g^{2/(4-D)}$ which is small if the 
coupling parameter $g$ is taken to be small). Thus,  
with the exponential accuracy the (Griffith-like)
contribution to the free energy coming from these local 
ferromagnetic states can be estimated by their
probability, eq.(\ref{gr4}):

\be
\lb{gr5}
F_{G} \; \sim \; \int_{0}^{\infty} d\xi 
         \int_{\R(\xi)}^{\infty} dL
         \exp\[ -\frac{1}{4u} (\t + \xi)^{2} \; L^{D} \] \;
\sim \; \int_{0}^{\infty} d\xi 
     \exp\[ -(const)\frac{(\t + \xi)^{2}}{u} \xi^{-D/2} \]
\ee
Here in the integration over $\xi$ the leading contribution
comes from the vicinity of the saddle-point value
 \be
\lb{gr6}
\xi_{*} \; = \; \fr{D}{4 - D} \; \t
\ee
(which is positive in dimensions $D<4$, 
and $\xi_{*} \gg \t_{g}$ provided $\t \gg \t_{g}$).
In this way we obtain
the following estimate for the off-perturbative
contributions coming from rare locally
ferromagnetic islands:
\be
\lb{gr7}
F_{G}  \; \sim \; \exp\[ -(const)\frac{\t^{(4-D)/2}}{u} \]
\ee
In fact, this result (including the value of the 
$(const)$ factor), as we will see in section IV,
 can be derived analytically 
in terms of the formal replica calculations
as the contribution from the  localized 
(instanton-like) solutions of the mean-field 
saddle-point equations \cite{grif-inst}.

\vspace{3mm}

In the presence of non-zero external magnetic field 
$h$ the situation becomes slightly more tricky.
Let us consider again the ferromagnetic island 
of the size $L$ with negative "local temperature" 
$(\t - \d \t) = -\xi$ which is described by 
the Hamiltonian,
eq.(\ref{gr2}). Note first of all, that the 
effective potential
\be
\lb{gr8}
U(\f) \; = \; -\2\xi\f^{2} + \4 g \f^{4} - h \f
\ee
has two minima only if the value of 
$\xi$ is not too small, namely at
\be
\lb{gr9}
\xi > \xi_{*}(h) \sim h^{2/3} g^{1/3},
\ee
otherwise, at $\xi < \xi_{*}(h)$, it has a unique
minimum (see Fig.1). In the case of the two minima, 
the state "up" ($\f > 0$) has the lower energy 
$E_{u}(\xi, h, L)$,
and it is the ground state of the island, 
while the state "down" ($\f < 0$) can be considered 
as the excitation since it has
higher energy, $E_{d}(\xi, h, L)$. 
The crucial point is that the excited state, 
$E_{d}$ disappears 
discontinuously at $\xi = \xi_{*}(h)$
such that the energy difference 
$\D E = E_{d} - E_{u}$ remains finite
at this critical value of $\xi$. 
Thus, the contribution to the total free energy
of such ferromagnetic islands can be estimated
as follows:

\begin{figure}[h]
\includegraphics[scale=0.35]{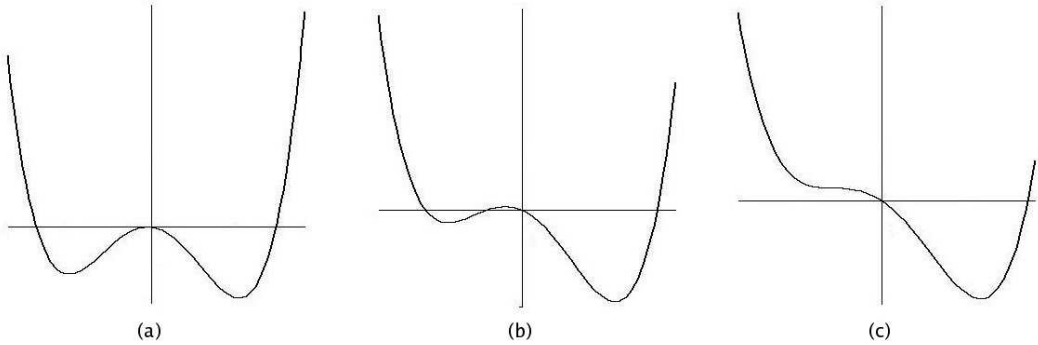}
\caption[]{Qualitative shape of the potential
$U(\f)$, eq.(\ref{gr8})):
(a) at $\xi \gg \xi_{*}(h)$;
(b) at $\xi > \xi_{*}(h)$; 
(c) at $\xi = \xi_{*}(h)$.}
  \label{fig1}
\end{figure}

\be
\lb{gr11}
F_{G}(h) \; \sim \; - \int_{\xi_{*}(h)}^{\infty} d\xi 
         \int_{\R(\xi)}^{\infty} dL \;
	 {\cal P}\[L, \xi\] \; \ln\[ \mbox{\Large $e$}^{-E_{u}} 
              + \mbox{\Large $e$}^{-E_{d}} \] \;
         - \;\int_{0}^{\xi_{*}(h)} d\xi 
         \int_{\R(\xi)}^{\infty} dL \;
	 {\cal P}\[L, \xi\] \; \ln\[ \mbox{\Large $e$}^{-E_{u}} \]
\ee 
where ${\cal P}[L, \xi]$ is the probability to have the
island of the size $L$ with the local temperature
$\xi$, eq.(\ref{gr4}). Since the integration over
$L$ sticks to the lower bound $\R$, we find:
\be
\lb{gr12}
F_{G}(h) \sim \int_{0}^{\infty} d\xi \;
	 P(\xi) \; E_{u}(\xi,h) \; - \;  
          \int_{\xi_{*}(h)}^{\infty} d\xi \; 
         P(\xi) \; \ln\[1 + e^{-\D E(\xi,h)} \]
\ee
where 
\be
\lb{gr13}
P(\xi) \; \sim \; 
   \exp\[ -(const)\frac{(\t + \xi)^{2}}{u} \xi^{-D/2} \]
\ee
On the other hand, similar considerations for the 
zero field case (when $E_{d} \equiv E_{u}$) yield

\be
\lb{gr14}
F_{G}(0) \; \sim \; -\int_{0}^{\infty} d\xi 
         \int_{\R(\xi)}^{\infty} dL \;
	 {\cal P}\[L, \xi\] \; 
	 \ln\[ 2 e^{-E_{u}(\xi,0)} \] \;
\; \sim \; \int_{0}^{\infty} d\xi \; 
	 P(\xi) \; \[ E_{u}(\xi,0)  - (\ln 2)\]  
\ee
Thus, for the free energy difference,
$\D F_{G}(h) = F_{G}(h) - F_{G}(0)$  we find:

\be
\lb{gr15}
\D F(h)  
\; \sim \; \int_{0}^{\infty} d\xi \; P(\xi) \;
      \[ E_{u}(\xi,h) - E_{u}(\xi,0)\]    
\; - \;\int_{\xi_{*}(h)}^{\infty} d\xi \; P(\xi) \; 
    \ln\[ \fr{1 + e^{-\D E(\xi,h)}}{2} \]
\; + \; (\ln 2) \int_{0}^{\xi_{*}(h)} d\xi \; P(\xi) 
\ee
In the limit $h \to 0$ the first two terms 
in the above equation provide regular functions
of $h$ (both $(E_{u}(\xi,h)-E_{u}(\xi,0))$ and
$\D E(\xi,h)$ go to zero as a power functions of
$h$ in the limit $h\to 0$),
while the last one is just the non-trivial
Griffiths contribution $\d F_{G} (h)$ 
which has the form 
of the essential singularity:

\be
\lb{gr16}
\d F_{G} (h) \sim (\ln 2) \int_{0}^{\xi_{*}(h)} d\xi \; P(\xi)
\sim \exp\( -(const) \frac{\t^{2}}{u}  \xi_{*}^{-D/2}(h) \)
\ee
Substituting here the value 
$\xi_{*}(h) = h^{2/3} g^{1/3}$,
eq.(\ref{gr9}), one eventually finds:

\be
\lb{gr18}
\d F_{G} (h) \; \sim  \exp\( -(const) \frac{\t^{2}}{u \; g^{D/6}}  
             \;  h^{-D/3}\)
\ee
In Section IV it will be demonstrated how this
result can be derived in terms of the
formal replica calculations. But first we have 
to formulate the general lines of the replica 
approach for the off-perturbative contributions.

\section{Off-perturbative replica calculations}

In this section I am going to formulate a general 
systematic approach for the calculations 
of off-perturbative contributions (if any) coming from
local minima states, which in the configurational space
are well separated from the ground state.

\begin{figure}[h]
\includegraphics[scale=0.50]{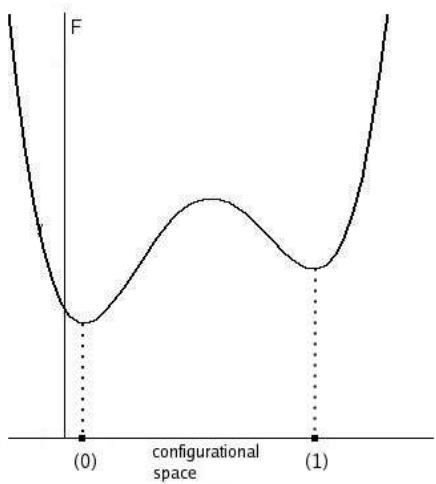}
\caption[]{Schematic structure of the free energy
landscape in the case of two well separated  
thermodynamically relevant valleys 
of the configurational space}
  \label{fig2}
\end{figure}

Let us consider a general random system 
described by a Hamiltonian $H\[\f(x)\]$, and
let us suppose that 
in addition to the
ground state, there is another thermodynamically 
relevant (Griffith) region of the 
configurational space located "far away"
from the ground state and separated from it 
by a finite barrier of the free energy (see Fig.2).
In other words, we suppose that the partition 
function (of a given sample) can be represented in
the form of two separate contributions:
\be
\lb{gr19}
Z  = \int {\cal D}\f(x) e^{-\b H} \; = \; 
  \mbox{\large $e$}^{-\b F_{0}} + \mbox{\large $e$}^{-\b F_{1}} 
 \; \equiv \;Z_{0} \; + \; Z_{1}
\ee
where $F_{0}$ is the contribution coming 
from the vicinity of the ground state,
and $F_{1}$  is the contribution of the 
Griffiths region. Then, for the averaged
over disorder total free energy we find:
\be
\lb{gr20}
{\cal F} = -\fr{1}{\b} \ol{\ln Z} 
= \ol{F_{0}} - \fr{1}{\b} 
\ol{\ln\[1 + Z_{1} Z_{0}^{-1}\]}
\ee
The second term here, which is just the Griffiths 
contribution, can be represented as follows:
\be
\lb{gr21}
F_{G} = - \fr{1}{\b} \sum_{m=1}^{\infty}
\fr{(-1)^{m-1}}{m} \ol{Z_{1}^{m} Z_{0}^{-m}}
= - \fr{1}{\b} \lim_{n\to 0} \sum_{m=1}^{\infty}
\fr{(-1)^{m-1}}{m} Z_{n}(m)
\ee	      
where
\be
\lb{gr22}
Z_{n}(m) = \prod_{b=1}^{m} \int {\cal D}\f^{(1)}_{b}
         \prod_{c=1}^{n-m} \int {\cal D}\f^{(0)}_{c} \; 
\mbox{\Large $e$}^{ -\b H_{n}\[\f^{(1)}_{1},...,\f^{(1)}_{m}, 
           \f^{(0)}_{1},...,\f^{(0)}_{n-m}\]}    
\ee
is the replica partition function
($H_{n}\[{\boldsymbol\phi}\]$ is the corresponding
replica Hamiltonian), 
in which the replica symmetry in the $n$-component 
vector field $\f_{a}$ ($a=1,...,n$) is assumed to be 
broken. Namely, it is supposed that the saddle-point
equations
\be
\lb{gr23}
\fr{\d H_{n}\[{\boldsymbol\phi}\]}{\d \f_{a}(x)} \; = \; 0
\;, \; \; \; \; (a = 1, ..., n)
\ee
have non-trivial solutions with the RSB structure
\be
\lb{gr24}
\f_{a}^{*}(x) = \left\{ \begin{array}{ll}
                 \f_{1}(x)   & \mbox{for $a = 1, ..., m$}
		 \\
		 \\
                 \f_{0}(x)   & \mbox{for $a = m+1, ..., n$}
                            \end{array}
                            \right.
\ee
with $\f_{1}(x) \not= \f_{0}(x)$, so that the integration
in the above partition function, eq.(\ref{gr22}), goes over
fluctuations in the vicinity of these solutions:
\bq
\lb{gr25}
\f^{(1)}_{b}(x) &=& \f_{1}(x) + \p_{b}(x) ,  \; \; \; \; 
          (b =  1, ..., m)
\nn
\f^{(0)}_{c}(x) &=& \f_{0}(x) + \h_{c}(x) , \; \; \; \; 
          (c =  1, ..., n-m)
\eq

It should be stressed that to be thermodynamically relevant,
the RSB saddle-point solutions, eq.(\ref{gr24}), 
should satisfy the following tree crucial conditions:

 (1) the solutions should be {\it local} in space, 
so that they are characterized by  {\it finite}
space sizes $R(m)$; in this case the partition 
function, eq.(\ref{gr22}), will be proportional to
the entropy factor $V/R^{D}(m)$ 
(where $V$ is the volume of the system), and
the corresponding free energy contribution 
$F_{G}$, eq.(\ref{gr21}), 
will be extensive quantity;
 
 (2) they should have {\it finite} energies 
$E(m) = H_{n}\[{\boldsymbol\f}^{*}\]$;

 (3) the corresponding Hessian matrix of these solutions 
should have all eigenvalues positive. 

Thus, in the systematic calculations one should
find all saddle-point RSB solutions $\f_{a}^{*}(x)$
(satisfying the above three requirements), 
eq.(\ref{gr24}), after that one has to compute
their energies $E(m)$ (for $n\to 0$), 
next one has to
integrate over the fluctuations in the vicinity of 
these solutions, and finally one has to 
sum up the series

\be
\lb{gr26}
F_{G} = -\fr{V}{\b} \sum_{m=1}^{\infty} \fr{(-1)^{m-1}}{m}  
   R^{-D}(m) \; \(\det \hat T\)^{-1/2}_{n=0} \; 
\mbox{\Large $e$}^{-\b E(m)}
\ee
where $\hat T$ is the ($n\times n$) matrix
\be
\lb{gr27}
T_{aa'} \; = \; 
\fr{\d^{2} H\[{\bf \f}\]}{\d\f_{a} \d\f_{a'}}\Bigl|_{\f=\f^{*}}
\ee

\vspace{5mm}

The above scheme of calculations can be easily 
generalized for an arbitrary number of the
Griffiths regions.
For example, let us
consider the situation which is qualitatively
represented in Fig.3, when in addition to the 
ground state, the system has {\it two} thermodynamically
relevant Griffiths states (which is just the case
for the considered random Ising model). 
In this case instead of eq.(\ref{gr19}) we will have
\begin{center}
\begin{figure}[h]
\includegraphics[scale=0.50]{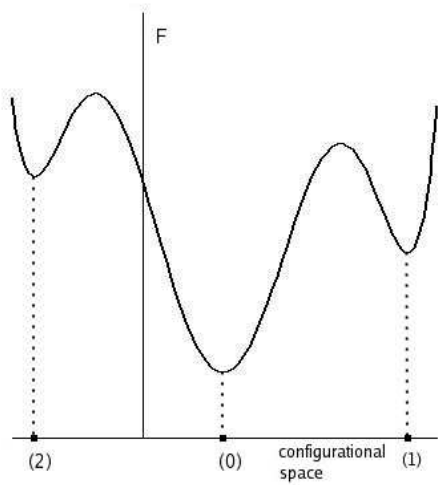}
\caption[]{Schematic structure of the free energy
landscape in the case of three well separated  
thermodynamically relevant valleys 
of the configurational space}
  \label{fig3}
\end{figure}
\end{center}

\be
\lb{gr28}
Z  = \int {\cal D}\f(x) e^{-\b H} \; = \; 
  \mbox{\large $e$}^{-\b F_{0}} + \mbox{\large $e$}^{-\b F_{1}} 
  + \mbox{\large $e$}^{-\b F_{2}}
  \; \equiv \; Z_{0} \; + \; Z_{1} \; + \; Z_{2}
\ee
and correspondingly, instead of eq.(\ref{gr21})
we find

\bq
\lb{gr29}
F_{G} &=& -\fr{1}{\b} 
\ol{\ln\[1 + Z_{1} Z_{0}^{-1} + Z_{2} Z_{0}^{-1}\]} \; = \;
- \; \fr{1}{\b} \sum_{m=1}^{\infty}
\fr{(-1)^{m-1}}{m} 
\sum_{k=0}^{m} C^{m}_{k} \; 
\ol{\(Z_{1}^{k} Z_{2}^{m-k}Z_{0}^{-m}\)}
\nn
\nn
&=& -\fr{1}{\b} \lim_{n\to 0} \sum_{m=1}^{\infty}
\fr{(-1)^{m-1}}{m} 
\sum_{k=0}^{m} C^{m}_{k} \; 
Z_{n}(k,m)
\eq	      
where $C^{m}_{k} = m!/(k!(m-k)!)$ is the combinatoric
factor.
Here, in the replica partition function

\be
\lb{gr30}
Z_{n}(k,m) = \prod_{b=1}^{k} \int {\cal D}\f^{(1)}_{b}
            \prod_{c=1}^{m-k} \int {\cal D}\f^{(2)}_{c}
            \prod_{d=1}^{n-m} \int {\cal D}\f^{(0)} \;    
	    \mbox{\Large $e$}^{-\b H_{n}
	    \[{\boldsymbol\f}^{(1)},
              {\boldsymbol\f}^{(2)},
	      {\boldsymbol\f}^{(0)}\]}
\ee
the integration is supposed to be performed  
in the vicinity of the saddle-point replica vector
\be
\lb{gr31}
\f_{a}^{*}(x) = \left\{ \begin{array}{ll}
                 \f_{1}(x),  & \mbox{for $a = 1, ..., k$}
		 \\
		 \\
                 \f_{2}(x),  & \mbox{for $a = k+1, ..., m$}
		 \\
		 \\
                 \f_{0}(x),  & \mbox{for $a = m+1, ..., n$}
                            \end{array}
                            \right.
\ee
(where $\f_{1}(x) \not= \f_{2}(x) \not= \f_{0}(x)$)
which is the solution of the saddle-point 
equations (\ref{gr23}).
Finally, for the Griffiths contribution, 
instead of eq.(\ref{gr26}) one obtain

\be
\lb{gr32}
F_{G} = -V \sum_{m=1}^{\infty} \fr{(-1)^{m-1}}{\b m}   
\sum_{k=0}^{m} C^{m}_{k} \; R^{-D} \(\det\hat T\)^{-1/2}_{n=0}
\mbox{\Large $e$}^{-\b E(k,m)}
\ee
where  
$E(k,m)= H_{n\to 0}\[{\boldsymbol\f}^{*}\]$ 
is the energy of a given solution,
eq.(\ref{gr31}), and $\hat T$ is the Hessian 
matrix, eq.(\ref{gr27}).

It is worthing to note that one can arrive 
to the same representations for the 
off-perturbative free energy contributions,
eqs.(\ref{gr26}) and (\ref{gr32}), 
in terms of the so called vector replica
symmetry breaking scheme\cite{vrsb,dos-book},
starting from 
the  standard replica approach for random systems
($F = -\b^{-1} \lim_{n\to 0} (\ol{Z^{n}} - 1)/n$).

In the next section we will implement
the programme described above for the 
concrete case of weakly disordered
ferromagnetic Ising model in the high temperature
paramagnetic phase.

\section{Replica instantons}

After the standard Gaussian averaging over random $\d\t(x)$ 
(described by the distribution, eq.(\ref{gr3})) of the $n$-th
power of the partition function one obtains the following 
replica Hamiltonian
\be
\lb{gr33}
H_{n}\[{\boldsymbol\f}\] \; = \;
 \I \Biggl[ \2 \sum_{a=1}^{n}\(\n\f_{a}\)^{2} + \2 \t \sum_{a=1}^{n}
 \f_{a}^{2} + 
 \4 g \sum_{a=1}^{n} \f_{a}^{4} -\4 u \sum_{a,b=1}^{n}  \f_{a}^{2} \f_{b}^{2}
 - h \sum_{a=1}^{n} \f_{a} \Biggr]
\ee
The corresponding saddle-point equations are
\be
\lb{gr34}
-\D\f_{a}(x) + \t \f_{a}(x) + g \f_{a}^{3}(x) 
-u \f_{a}(x) \(\sum_{b=1}^{n} \f_{b}^{2}(x)\) = h
\ee  
Substituting here the anzats, eq.(\ref{gr31}),
the above
saddle-point equations are reduced to
\be
\lb{gr35}
-\D\f_{i} + \t \f_{i} + g \f_{i}^{3} - u \f_{i} S = h
\ee
($i=1,2,0$) where
\be
\lb{gr36a}
S \equiv \sum_{a=1}^{n} \f_{a}^{2}(x) \; = \;  
k\f_{1}^{2} + (m-k)\f_{2}^{2} + (n-m)\f_{0}^{2}
\ee
which in the limit $n\to 0$ turns into
\be
\lb{gr36}
S \; = \; k\f_{1}^{2} + (m-k)\f_{2}^{2} - m\f_{0}^{2}		
\ee
Substituting  eqs.(\ref{gr31}), 
into the Hamiltonian, eq.(\ref{gr33}), 
for the energy of this configuration
(at $n\to 0$) we obtain:

\be
\lb{gr37}
E(k,m) = \I \Biggl[ \fr{k}{2} \(\n\f_{1}\)^{2} + 
     \fr{(m-k)}{2} \(\n\f_{2}\)^{2} -
  \fr{m}{2} \(\n\f_{0}\)^{2} +
        U(\f_{1},\f_{2},\f_{0}) \Biggr]
\ee
where
\bq
\lb{gr37a}
U(\f_{1},\f_{2},\f_{0}) &=& 	
 \2 \t \[k\f_{1}^{2}+(m-k)\f_{2}^{2}-m\f_{0}^{2}\] 
\; + \; \4 g \[k\f_{1}^{4}+(m-k)\f_{2}^{4}-m\f_{0}^{4}\]               
\nn
\nn
&-& \4 u \[k\f_{1}^{2}+(m-k)\f_{2}^{2}-m\f_{0}^{2}\]^{2}
\; - \; h \[k\f_{1}+(m-k)\f_{2}-m\f_{0}\] 
\eq
and the fields $\f_{1}(x), \f_{2}(x)$ and $\f_{0}(x)$
are defined by the equations (\ref{gr35}) and (\ref{gr36}).

\subsection{Zero external magnetic field}

First of all, we note that
at $h=0$, due to the symmetry $\f \to -\f$ 
the solution of eqs.(\ref{gr35})-(\ref{gr36}) takes the form:
\bq
\lb{gr38}
\f_{1}(x) = -\f_{2}(x) &\equiv& \f(x) 
\nn
         \f_{0}(x) &=& 0
\eq	 
where the function $\f(x)$ is 
defined by the equation
\be
\lb{gr39}
-\D\f(x) + \t \f(x) - \lm(m) \f(x)^{3} = 0
\ee
which is controlled by the parameter
\be
\lb{gr40}
\lm(m) = u m  - g
\ee
In what follows this parameter will be assumed 
to be {\it positive}.
In other words, the solution, which we are going 
to derived below,
exists only for $m$ such that 
\be
\lb{gr41}
m \; > \; \[\fr{g}{u}\]
\ee
Substituting eqs.(\ref{gr38}) into eq.(\ref{gr37})-(\ref{gr37a}) 
for the energy
of this solution we obtain

\be
\lb{gr42}
E(k,m) \equiv E(m) = m \I \Biggl[\2 (\n\f)^{2}  +
                           \2 \t \f^{2} - \4 \lm(m) \f^{4} \Biggr] 
\ee
Note here, that one should not be confused by the "wrong" 
sign of the
$\f^{4}$ term in the above expression (which for the usual
field theory would indicate its absolute instability).
In fact, as we will see below, the integration over the 
replica fluctuations around
considered solution in the limit $n\to 0$ yields the Hessian 
matrix which has all the eigenvalues positive
(this is quite standard situation for the replica theory:
in the limit $n\to 0$, when the number of certain
parameters become negative,   everything turns
"up down", so that minima of the physical quantities 
turns into maxima of the corresponding replica 
quantities\cite{MPV-book,dos-book}).

Rescaling the fields,
\be
\lb{gr43}
\f(x) = \sqrt{\fr{\t}{\lm(m)}} \psi(x/R_{c}(\t))
\ee
(where $R_{c}(\t) = \t^{-1/2}$),
instead of eq.(\ref{gr39}) one get the differential 
equation which contains no parameters:
\be
\lb{gr44}
-\D \psi(z) + \psi(z) - \psi^{3}(z) = 0
\ee
Correspondingly, for the energy, eq.(\ref{gr42}), one obtains:
\be
\lb{gr45}
E(m) = \fr{m}{um - g} \t^{(4-D)/2} E_{0}
\ee
where
\be
\lb{gr46}
E_{0} = \int d^{D}z \[ \2 (\n\psi(z))^{2} + \2 \psi^{2}(z) - \4 \psi^{4}(z) \]
\ee
The equation (\ref{gr44}) is well know in the field theory 
(see e.g. Ref.[\onlinecite{inst}]):
in dimensions $1< D < 4$  it has the smooth (with $\psi'(0) =0$)
spherically symmetric instanton-like solution such that:
\bq
\lb{gr47}
\psi(z \leq 1) &\sim& \psi(0) \sim 1, 
\nn
\nn
\psi(z \gg  1) &\sim& \mbox{\Large $e$}^{-z} \to 0. 
\eq
The energy $E_{0}$, eq.(\ref{gr46}), 
of this  solution is a finite and {\it positive} number.
In particular,
at $D = 3$, $\psi_{0} \simeq 4.34$ and $E_{0} \simeq 18.90$.
Note that according to the rescaling, eq.(\ref{gr43}), 
the size of this instanton solution 
in terms of the original fields $\f(x)$
is $R_{c} = \t^{-1/2}$. This size does not depends on $k$ and $m$,
and it coincides with the usual mean-field correlation 
length of the Ginsburg-Landau theory. 
Note also that due to the symmetry $\f \to -\f$ 
of the considered solution,
its parameters do not depend on $k$. Thus, we can perform
the summation over $k$ in the series, eq.(\ref{gr32}), 
which yields the trivial factor $2^{m}$:
\be 
\lb{gr48}
 F_{G}  \; \simeq \; 
- V \R(\t)^{-D} \sum_{m > [g/u]}^{\infty} \fr{(-1)^{m-1}}{m} 2^{m} 
\(\det\hat T\)^{-1/2}_{n=0} \;  
\exp\(-E_{0} \fr{m}{um-g} \t^{(4-D)/2} \)
\ee
In other words, the considered two-step 
structure, eq.(\ref{gr31})-(\ref{gr32}), is equivalent
to the one-step anzats, eq.(\ref{gr24}),
\be
\lb{gr49}
\f_{a}^{*}(x) = \left\{ \begin{array}{ll}
                 \sqrt{\fr{\t}{\lm(m)}} \psi(x\sqrt{\t}) 
		  & \mbox{for $a = 1, ..., m$}
		 \\
		 \\
                 0  & \mbox{for $a = m+1, ..., n$}
                            \end{array}
                            \right.
\ee
which has additional degeneracy factor $2^{m}$.

The final step is the integration over fluctuations,
which define the Hessian factor $(\det\hat T)$.
Introducing small fluctuations $\p_{a}(x)$ near the above
instanton solution, $\f_{a}(x) = \f^{*}_{a}(x) + \p_{a}(x)$, 
in the Gaussian approximation we get the following 
Hamiltonian for the fluctuating fields:
\be
\lb{gr50}
H[\p] \simeq  \I \[ \2 \sum_{a=1}^{n} \(\n\p_{a}\)^{2} + 
            \2 \t \sum_{a,a'=1}^{n} K_{aa'}(x) \p_{a}\p_{a'}  \] 
\ee
where the matrix $K_{aa'}(x)$ contains  the $m\times m$ block:
\be
\lb{gr51}
K^{(m)}_{bb'}(x)  = 
\(1 - \fr{um - 3g}{um - g} \psi^{2}(x\sqrt{\t})\)\d_{bb'} - 
         \fr{2u}{um-g} \psi^{2}(x\sqrt{\t})
\ee
($b,b' =1, ..., m$) and the diagonal elements 
for the remaining $(n-m)$ replicas:
\be
\lb{gr52}
K^{(n-m)}_{cc'} = 
\(1 - \fr{um}{um-g} \psi^{2}(x\sqrt{\t})\) \d_{cc'}
\ee
($c,c' = m+1, ..., n$). 
To obtain the explicit result for the Hessian, let us 
approximate the instanton solution, eq.(\ref{gr47}), 
by the $\theta$-like function
\be
\lb{gr53}
\psi(z) = \left\{ \begin{array}{ll}
                 \psi_{0},  & \mbox{for $0\leq z \leq 1$}
		 \\
		 \\
                 0 ,     & \mbox{for $z > 1$}
	    \end{array}
                            \right.
\ee
Then, the approximate Hamiltonian for the
fluctuating fields takes much more simple form 

\be
\lb{gr54}
H[\p] \; \simeq \; \2 \sum_{a,a'=1}^{n} 
\int_{|p|\gg\sqrt{\t}} \fr{d^{D}p}{(2\pi)^{D}} 
      \[ p^{2} \d_{aa'} + \t K_{aa'} \] \p_{a}(p)\p_{a'}(-p) 
\; + \; \2 \sum_{a=1}^{n} \int_{|p|\ll\sqrt{\t}} 
\fr{d^{D}p}{(2\pi)^{D}}  p^{2} |\p_{a}(p)|^{2}
\ee
Here the $p$-independent matrix $K_{aa'}$ is given 
by eqs.(\ref{gr51})-(\ref{gr52}), where
instead of the function $\psi(x\sqrt{\t})$ one has 
to substitute the constant $\psi_{o}$.

The integration over the modes with 
momenta $p \ll \sqrt{\t}$ 
(corresponding to the scales much bigger 
than the size of the instanton), which are  
described by the second term of the above Hamiltonian, 
gives the contribution
of the form $\exp(- n V f_{RS} )$. This contribution
is irrelevant in the limit $n\to 0$, which 
is quite natural,
because all such contributions must be already
contained in the perturbative part of the free energy
$F_{0}$, eq.(\ref{gr20}), which we do not consider here.

The integration over the modes with momenta 
$p \gg \sqrt{\t}$ is slightly
cumbersome but straightforward:

\be
\lb{gr55}
\(\det\hat T\)^{-1/2} 
\simeq \exp\[ -\fr{\t^{-D/2}}{2} \int_{p \gg \sqrt{\t}} 
\fr{d^{D}p}{(2\pi)^{D}} \; \Tr \; \ln \(p^{2}\d_{aa'} +\t K_{aa'}\) \]
\ee
The matrix under the logarithm in the above equation 
contains $(m-1)$ eigenvalues:
\be
\lb{gr56}
\lm_{1} = p^{2} + \t \(1 - \fr{um - 3g}{um - g} \psi_{o}^{2}\)
\ee
one eigenvalue:
\be
\lb{gr57}
\lm_{2} = p^{2} + \t \(1 - \fr{um - 3g}{um - g} \psi_{o}^{2}\) 
- \t \fr{2um}{um-g} \psi_{o}^{2}
\ee
and $(n-m)$ eigenvalues:
\be
\lb{gr58}
\lm_{3} = p^{2} + \t \(1 - \fr{um}{um - g} \psi_{o}^{2} \)
\ee
which are all positive.
Substituting these eigenvalues into eq.(\ref{gr55}), 
after simple algebra 
in the limit $n\to 0$ one eventually obtains the 
following result:
\be
\lb{gr59}
\(\det\hat T\)^{-1/2}  \simeq 
\exp\[  \fr{3m}{2(um-g)} g \psi_{o}^{2}\]
\ee
Finally, substituting this value into eq.(\ref{gr48})
we find 

\be 
\lb{gr60}
F_{G}  \; \simeq \;
- V \R(\t)^{-D} \sum_{m > [g/u]}^{\infty} \fr{(-1)^{m-1}}{m} \; 2^{m}  \;
\exp\[-E_{0} \fr{m}{um-g} \t^{(4-D)/2} +
    \fr{3m}{2(um-g)} g \psi_{o}^{2}\]
\ee
Here one can note that under condition
\be
\lb{gr61}
\t \gg \t_{g} = g^{2/(4-D)}
\ee
the second term in the exponential of eq.(\ref{gr60})
(which is the fluctuations contribution)
can be neglected compared to the first one.
This is not surprising because eq.(\ref{gr61})
is nothing else, but the good old Ginzburg-Landau condition
which defines the temperature region away from $\T$,
where the critical fluctuations are irrelevant,
and the behavior of the system is described
by the mean-field exponents.

The exact summation of the series in eq.(\ref{gr60})
is rather tricky problem, but with the exponential accuracy
it can be estimated in a very simple way.
One can easily see that in the limit of weak disorder,
at $u \ll g$, the leading contribution in this summation
comes from the region $m \gg g/u \gg 1$ 
(where the exponential
factor in eq.(\ref{gr60}) becomes $m$-independent)
and this contribution is
\be
\lb{gr62}
 F_{G} \sim \exp\(-E_{0} \fr{\t^{(4-D)/2}}{u} \)
\ee
We see that 
this result nicely coincides with the 
naive "hand-waving" estimate, eq.(\ref{gr7}), where
the $(const)$ factor is the instanton energy $E_{0}$
(in three dimensions $E_{0} \simeq 18.9$). 

\subsection{Non-zero external magnetic field}

Technically, the situation in non-zero magnetic field
becomes much more cumbersome, but on a qualitative level,
the main idea of the approach remains very simple.
According to the physical discussion of Section II,
non-analytic contribution to the free energy in
the presence of external magnetic field appears
due to the fact, that some of the instanton-like
configurations (of the type, considered in the previous
subsection) disappear via a finite jump. 

To understand that, let us consider first the structure
of the potential energy 
$U(\f_{1},\f_{2},\f_{0})$, eq.(\ref{gr37a}).
The extrema of this potential are defined by the 
three equations 
\be
\lb{gr64}
\t \f_{i} + g \f_{i}^{3} - u \f_{i} S = h
\ee
($i=1,2,0$), where 
\be
\lb{gr65}
S \; = \; k\f_{1}^{2} + (m-k)\f_{2}^{2} - m\f_{0}^{2}		
\ee
Simple algebraic manipulations 
reduce these equations to
\be
\lb{gr66}
\f_{1} + \f_{2} + \f_{0} = 0
\ee
\be
\lb{gr67}
\f_{1} \f_{2} (\f_{1} + \f_{2}) = -\fr{h}{g}
\ee
and
\be
\lb{gr68}
\t + \[g + u(m-k)\] \f_{1}^{2} + \(g + uk\) \f_{2}^{2} + 
    \(g + 2 u m \) \f_{1} \f_{2} = 0
\ee
First, treating the magnetic field $h$ here as a small correction
to the zero-field solution, 
\bq
\lb{gr69}
\f_{1}^{(h=0)} = -\f_{2}^{(h=0)} &\equiv& \f(m) = \sqrt{\fr{\t}{um - g}}
\nn
\nn
        \f_{0}^{(h=0)} &=& 0
\eq
one easily finds
\bq
\lb{gr70}
\f_{1} &\simeq& \f(m) + \fr{h}{g \f^{2}(m)} \; + \; O\(h^{2}\) ,
\nn
\nn
\f_{2} &\simeq& -\f(m) \; + \; O\(h^{2}\),
\nn
\nn
\f_{0} &\simeq& -\fr{h}{g \f^{2}(m)} \; + \; O\(h^{2}\)
\eq
It is clear that this linear in $h$ shift of the 
extremum of the potential $U(\f_{1},\f_{2},\f_{3})$
provide not more than a linear in $h$ corrections to the 
zero field instanton space configuration 
as well as to its free energy contribution 
considered in the previous subsection.
It has to be noted,
however, that the above result, eq.(\ref{gr70}),
is valid only until the summation parameters
$k$ and  $m$
are not too large:
\be
\lb{gr71}
m, k \; \ll \;  \fr{\t}{u} \(\fr{g}{h}\)^{2/3}
\ee
Simple analysis shows that as 
$k$ and $m$ grow,
the values of (negative) $\f_{2}(k,m)$ and $\f_{0}(k,m)$ become 
closer and closer to each other, 
and finally, one arrive to the critical configuration
when their values coincide. Substituting  
$\f_{0} = \f_{2} \equiv \f_{2}^{(cr)}$
into equations (\ref{gr66})-(\ref{gr68}) one easily finds that
\bq
\lb{gr72}
\f_{2}^{(cr)} &=& - \(\fr{h}{2g}\)^{1/3}
\nn
\nn
\f_{1}^{(cr)} &=& 2 \(\fr{h}{2g}\)^{1/3}
\eq
and this critical configuration (in the limit $h\to 0$)
takes place at
\be
\lb{gr73}
k_{c}(h) \; \simeq \; \fr{\t}{3u} \(\fr{2g}{h}\)^{2/3}
\ee
for arbitrary value of $m \geq k_{c}$. 
At larger values of $k$ the system of equations
(\ref{gr66})-(\ref{gr68}) have no solutions at all.
The simplest way to understand this, is to consider 
the eqs.(\ref{gr67}) and (\ref{gr68}) for
$k$ and $m$  much bigger than $k_{c}$. 
In this case eq.(\ref{gr68}) would require
the values of $\f_{1}$ and $\f_{2}$ 
as the functions of $k$ and $m$ to be of order $k^{-1/2}$
and $m^{-1/2}$, 
tending to zero as $m, k \to\infty$, while eq.(\ref{gr67})
tells that both $\f_{1}$ and $\f_{2}$
must remain finite.
Thus, the summations in the instanton free
energy contribution, eq.(\ref{gr32}), has to be 
limited 
by  the finite value $k_{c}(h)$: 

\be
\lb{gr74}
F_{G}(h) = -V \sum_{m=m_{0}}^{k_{c}} \fr{(-1)^{m-1}}{m}   
\sum_{k=0}^{m} C^{m}_{k}  
\fr{\mbox{\Large $e$}^{-E(k,m;h)}}{R^{D}
\(\det\hat T\)^{1/2}_{n=0}}
\;
- V \sum_{m=k_{c}}^{\infty} \fr{(-1)^{m-1}}{m}   
\sum_{k=0}^{k_{c}} C^{m}_{k}  
\fr{\mbox{\Large $e$}^{-E(k,m;h)}}{R^{D}
\(\det\hat T\)^{1/2}_{n=0}}
\ee
(here $m_{0} = \[g/u\]+1$).
This expression has to be compared with the 
corresponding summation at $h=0$, studied in the 
previous subsection. Although all the terms
in these series are non-analytic functions 
of the disorder parameter $u$ 
(see previous subsection),
until $k, m  \ll k_{c}(h)$, the difference 
between the terms 
with $h\not=0$ and the corresponding terms 
with $h=0$
can be represented in the form of corrections 
in powers of
$h$. On the other hand, as we will see below,
the terms with 
$k \sim k_{cr}(h)$ are non-analytic functions of 
$h$, and their differences with
the corresponding zero-field contributions  
(in particular the differences of the instanton
energies) can not be expended in powers 
of $h$. It is these terms, which are the 
contributions of the critical instantons 
which yield the non-analytic in $h$ part
of the free energy $\d F_{G}(h)$. Rigorous extracting of this 
piece must be a very difficult problem (it would
require derivation of the instantons energy
for arbitrary $k$, $m$ and $h$, as well as 
summations of the full series in eq.(\ref{gr74})),
but with the exponential accuracy the form
of this non-analytic singularity is defined 
only by the energy $E^{(cr)}(h)$ of the critical instanton,
\be
\lb{gr74a} 
\d F_{G}(h) \; \sim \; \exp\[ -E^{(cr)}(h)\]
\ee
Let us study the instanton configuration 
and estimate its energy in the vicinity
of its critical state. For that let us rescale
the fields, 
\be
\lb{gr75}
\f_{i} = \(\fr{h}{g}\)^{1/3} \psi_{i}\(x/\R\)
\ee
$(i=1,2,0)$ where 
\be
\lb{gr76}
\R = h^{-1/3} g^{-1/6}
\ee
defines the spacial size of the critical
instanton. In terms of the rescaled fields
$\psi$, the energy of the instanton takes the
form
\be
\lb{gr77}
E(k,m) \; = \; \fr{\R^{D} h^{4/3}}{ g^{1/3}}  
\int d^{D}z \Biggl[ \fr{k}{2} \(\n\psi_{1}\)^{2} +
        \fr{(m-k)}{2} \(\n\psi_{2}\)^{2} -
        \fr{m}{2} \(\n\psi_{0}\)^{2} +
	U\(\psi_{1},\psi_{2},\psi_{0}\) \Biggr]
\ee
where
\bq
\lb{gr78}
 U\(\psi_{1},\psi_{2},\psi_{0}\) &=&
        \2 \tl{\t}\[k\psi_{1}^{2}+(m-k)\psi_{2}^{2}-m\psi_{0}^{2}\] 
\; - \; \4 \tl{u} \[k\psi_{1}^{2}+(m-k)\psi_{2}^{2}-m\psi_{0}^{2}\]^{2}
\nn
\nn
&+&           \4 \[k\psi_{1}^{4}+(m-k)\psi_{2}^{4}-m\psi_{0}^{4}\]               
\; - \;         \[k\psi_{1}+(m-k)\psi_{2}-m\psi_{0}\] 
\eq
is the potential controlled by two rescaled
parameters
\be
\lb{gr79}
\tl{\t} \; = \; \fr{\t}{g^{1/3} \; h^{2/3}} \; ; \; \; \; \; \; \;
\tl{u} = \fr{u}{g}
\ee
The instanton configuration is defined by 
the three equations,
\be
\lb{gr80}
-\D\psi_{i} + \tl{\t} \psi_{i} + \psi_{i}^{3} - \tl{u} \psi_{i} S = 1
\ee
($i=1,2,0$) where
\be
\lb{gr81}
S \; = \; k\psi_{1}^{2} + (m-k)\psi_{2}^{2} - m\psi_{0}^{2}		
\ee
Let us consider these equations in the limit
$h \to 0$ and and for the values of $k$ and $m$ 
of order of $k_{c} \sim h^{-2/3} \to \infty$.
There are two types of terms in eqs.(\ref{gr80}):
(1) $\tl{\t} \psi_{i}$ and $\tl{u} \psi_{i} S$ which are 
both of order $h^{-2/3} \to \infty$; and (2),
the rest of the terms, which are of order of one.
In the limit $h \to 0$ the diverging terms 
must balance each other:
\be
\lb{gr82}
\tl{\t} \psi_{i} \sim \tl{u} 
\(k\psi_{1}^{2} + (m-k)\psi_{2}^{2} - m\psi_{3}^{2}\) \psi_{i} 
\ee
This condition is consistent with the requirement
that both $k$ and $m$
are of the order of $k_{c}(h)$, eq.(\ref{gr73}).
Substituting these estimates into eqs.(\ref{gr77}),
(\ref{gr78}), for the energy of the instanton
in its critical configuration we get
\be
\lb{gr83}
E^{(cr)} \; \sim \; \fr{\t^{2}}{u g^{D/6}} h^{-D/3}
\ee
Thus, with the exponential accuracy, 
the field dependent non-analytic part of the 
free energy has the following explicit form

\be
\lb{gr84}
\d F_{G}(h) \sim \mbox{\large $e$}^{-E^{(cr)}} \sim
\exp\(-(const) \fr{\t^{2}}{u g^{D/6}} h^{-D/3} \)
\ee
which perfectly agree with the mean-field "hand-waving"
estimate, eq.(\ref{gr18}). It has to be noted, however,
that in deriving the above result we have
neglected the effects of the critical fluctuations,
which is justified only if we consider the system
at temperatures not too close to the critical point,
$\t \gg \t_{g} = \t^{2/(4-D)}$. In the close vicinity
of $\T$, at $\t \ll \t_{g}$, the situation becomes 
much more complicated: here one would have 
to combine the systematic (renormalization-group) 
integration over fluctuation with the background
instanton solutions.

\section{Conclusions}

In this paper the systematic approach for the
off-perturbative calculations in disordered systems
has been formulated (Section III). 
It has been demonstrated how the proposed scheme
works in the most simple but non-trivial case
of weakly disordered ferromagnetic Ising model
away from its critical region. Here the non-analytic
(as the functions of the disorder parameter 
and the external magnetic field) 
contributions to the free energy has been derived,
eqs.(\ref{gr62}), (\ref{gr84}), and it has been demonstrated
that in terms of the replica field theory 
such contributions appear due to instanton-like
excitations. Of course, it is hardly possible to
register  the presence of these
exponentially small parts of the free energy
both in real and in numerical experiments,
and in this sense the present results 
has mostly pure theoretical interest.
On the other hand, thinking about the others
much more complicated problems of the statistical
mechanics of disordered systems, the investigations
made in this paper look rather promising.
In particular, it does not look completely 
unrealistic to try to combine the present 
off-perturbative scheme with the renormalization-group
treatment of the critical fluctuations,
to settle down recent suspects \cite{critical-rsb}
that off-perturbative degrees of freedom could be quite 
relevant in the vicinity of the critical point, so that 
the nature of the phase transition in 
random ferromagnetic systems may appear to be not as
simple as it was thought in early days of the 
theory of the critical phenomena in disordered materials.
\cite{RG-disorder}.

\end{document}